\documentclass[american, 12pt, a4paper, titlepage]{article}
\usepackage[T1]{fontenc}
\usepackage[latin1]{inputenc}
\usepackage{lmodern}
\usepackage{graphicx}
\usepackage{booktabs}
\usepackage[chatter]{rotating}
\usepackage{fancyvrb}
\usepackage{amssymb,amsmath}
\usepackage{float}
\usepackage{tikz}
\usetikzlibrary{arrows,positioning,shadows,shapes}
\usepackage{subfigure}
\usepackage{stmaryrd}
\usepackage{cite}
\usepackage{xcolor}
\usepackage{amsfonts}
\usepackage{url}
\usepackage{pdfpages}
\usepackage{colortbl}
\usepackage{xcolor}
\usepackage{multirow}
\usepackage{textcomp}
\usepackage{hanging}
\usepackage{ulem}
\usepackage{booktabs}
\usepackage{algorithm,algorithmic}

\begin{document}

\title{{\bf Re}peated undersampling in {\bf PrInDT (RePrInDT)}: \\Variation in undersampling and prediction, and ranking of predictors in ensembles}

\author{
Claus Weihs\\
TU Dortmund University\\
Faculty of Statistics\\
{\normalsize claus.weihs@tu-dortmund.de}
\and
Sarah Buschfeld\\
TU Dortmund University\\
Faculty of Cultural Studies\\
{\normalsize sarah.buschfeld@tu-dortmund.de}
}
\date{\vspace{1cm}\today}

\begin{titlepage}

\maketitle

\end{titlepage}

\noindent{\bf Abstract}\vspace{0.2cm}\\
In this paper, we extend our PrInDT method (Weihs \& Buschfeld 2021a) towards undersampling with different percentages of the smaller and the larger classes (psmall and plarge), stratification of predictors, varying the prediction threshold, and measuring variable importance in ensembles. An application of these methods to a linguistic example suggests the following: 1. In undersampling, a careful selection of the percentages plarge and psmall is important for building models with high balanced accuracies; 2. Stratification of predictors does not majorly enhance balanced accuracies; 3. Lowering the prediction threshold for the smaller class turns out to be an alternative method to undersampling because it increases the likelihood of the smaller class being selected. Finally, we introduce a method for ranking predictor importance that allows for a straightforward interpretation of the results.

\section{Introduction}

Building on our two previous studies of children's English in Singapore and England, PrInDT (Weihs \& Buschfeld 2021a) and NesPrInDT (Weihs \& Buschfeld 2021b), we further elaborate on the problem of how to best handle unbalanced data sets in linguistics.

In Weihs \& Buschfeld (2021a), we undersampled only the large class as we wanted to keep the new method as simple as possible. In the present study, we extend the undersampling method to the smaller class (Section~\ref{subsec:4.1}). In addition, when applying PrInDT, we also realized that some of the predictors come with very different frequencies of factor levels (e.g. more tokens produced by female than by male participants). In such cases, the proportion of frequencies between the two classes is typically not retained in random undersampling. In the present paper, we therefore stratify the individual predictor levels so that each class is undersampled in a way that the percentages of the predictor levels from the original data set are retained (Section~\ref{subsec:4.1}). Alternatively, we suggest a stratification method which ensures that each predictor value is represented by a minimum number of tokens (Section~\ref{subsec:4.1}). We further saw in the PrInDT paper that sometimes the frequencies of the two classes in a terminal node of a conditional inference tree (ctrees; Hothorn et al. 2006) may be nearly equal. In such cases, class prediction is ambiguous. In Section~\ref{subsec:4.2}, we therefore implement a method that allows for adjusting the threshold for choosing the smaller class for prediction. Finally, we introduce a method for ranking predictors in PrInDT ensembles, similar to the ranking of predictors in Random Forests, which is a widely used method in linguistic studies (Section~\ref{subsec:4.3}).

We illustrate the PrInDT extensions introduced in this paper on the basis of the same linguistic example as in the PrInDT paper (Weihs \& Buschfeld 2021a). We discuss the data set and linguistic objectives in Section~\ref{sec:2}. In Section~\ref{sec:3}, we briefly summarize the PrInDT approach. In Section~\ref{sec:4}, we introduce its extensions and in Section~\ref{sec:5} we apply them to the linguistic example. A conclusion is provided in Section~\ref{sec:6}.

\section{Linguistic objectives and data}\label{sec:2}

In the PrInDT paper, we statistically modeled the influence of intra- and extralinguistic variables on the acquisition of English subject pronouns by Singaporean and British children (for details, see Weihs \& Buschfeld 2021a). The data were collected by one of the authors and elicited systematically in video-recorded task-directed dialogue between the researcher and the children. The data collection procedure contained several parts: a grammar elicitation task, a story retelling task, elicited narratives, and free interaction. The recorded material was orthographically transcribed and manually coded for the realization of subject pronouns (realized vs. zero = not realized). The aim of our analysis was to find prediction rules for the use of subject pronouns, which were either realized by the children, or zero. Examples 1 and 2 illustrate the use of zero pronouns as produced by the Singaporean children:
\begin{itemize}
\item[1.] Researcher: [. . . ] what do you do with your friends? Do you play with them?\\
Child: [ ø I] Play with them. Sometimes drawing. [. . . ]\\
Child: Sometimes [ ø WE] play some fun things.
\item[2.] Child: I think in MH370, I think they can find because [ ø IT] is easy to go there [... ].
\end{itemize}

In order to predict the realization of subject pronouns, we modeled their dependence on the intralinguistic variable \textsc{pronoun} (PRN) and a number of extralinguistic variables, i.e. \textsc{ethnicity} (ETH), \textsc{age} (AGE), \textsc{sex} (SEX), \textsc{linguistic background} (LiBa), and \textsc{mean length of utterance} (MLU) (for details see Weihs \& Buschfeld 2021b). We thereby determined whether any of these variables had a statistically significant influence on the results. All in all, we extracted 6,146 tokens of the values of the subject pronoun variable, each with the full set of extra- and intra-linguistic variables. 528 of these were realized as zero pronouns.

\section{The PrInDT approach}\label{sec:3}

The methodological approach developed and pursued in this paper is an extension of the PrInDT approach (Weihs \& Buschfeld 2021a), which is based on decision trees (ctrees, Hothorn et al. 2006). Let us briefly recall the basic terminology and underlying ideas of this approach, in particular the notions of resampling, undersampling, prediction, and balanced accuracy. In PrInDT, predictive power is identified as the most important criterion to assess decision trees. In order to assess predictive power in data with a high imbalance between the small and the large classes, a simple way of undersampling is used as a resampling procedure. We repeatedly employed the full sample of the smaller class together with a small percentage of the larger class as the training set of the decision tree. The predictive power was assessed by means of the balanced accuracy of the two classes on the full sample. This way, accuracies of decision trees from different undersamples can be easily compared. As an additional criterion, the interpretability of the trees is considered in PrInDT, i.e. uninterpretable combinations of variable values are automatically excluded from the trees. Furthermore, we combined interpretable trees with high predictive power to ensembles in order to include more than one undersample in the process of interpretation and prediction. For further details on the PrInDT approach, the interested reader is referred to Weihs \& Buschfeld (2021a).

\section{Extensions of PrInDT}\label{sec:4}

\subsection{Undersampling}\label{subsec:4.1}

In PrInDT, we undersampled only the larger class and always included the full sample of the smaller class when constructing the model to keep the method as simple as possible. Therefore, no token of the smaller sample was held out for prediction. Thus, in the balanced accuracy, the accuracy of the smaller class only represented the fit quality of the model. The idea behind this approach was that we did not want to lose any of the observations of the smaller class for model construction because of their low frequency. Still, the balanced accuracy of a model should represent its predictive power. In the present paper, we therefore aim to assess the effect of \textbf{undersampling of the smaller class} on the predictive power of the overall model. To this end, we extended the PrInDT procedure by allowing undersampling of the smaller class as well. Due to the high imbalance in the sample size of the smaller and larger classes, the undersampling percentage of the larger class needs to be lower than the undersampling percentage of the smaller class. We therefore suggest taking different percentages (plarge and psmall) of undersampling for the two classes.

In addition, when applying PrInDT, we also realized that some of the predictors come with very different frequencies of factor levels (e.g. more tokens produced by female children than male children). In such cases, the proportion of the frequencies of the two classes is typically not retained in random undersampling. This is particularly problematic if the smaller class is not adequately represented in the resampled training sets. In the present paper, we therefore stratify the individual predictor levels so that each class is undersampled in a way that the percentages of the predictor levels from the original data set are retained ({\bf proportional stratification, prop.strat.}). This is one way to guarantee that even rare levels are adequately represented in the undersamples.

Alternatively, we suggest a stratification method which ensures that each predictor value is represented by a priorily specified minimum number of tokens ({\bf minimum criterion stratification, min.crit.strat.}), depending on the study. This guarantees that each group of predictor values is represented in each undersample without intervening too strongly into the random selection of values. If the priorily specified minimum is not reached by random undersampling, the procedure is repeated up to 10 times. If the set minimum can still not be reached, it is not compatible with plarge and psmall and needs to be redefined.

\subsection{Prediction threshold}\label{subsec:4.2}

\begin{figure}[H]
\centering
\vspace{-0.9cm}
\includegraphics[width=1.0\textwidth]{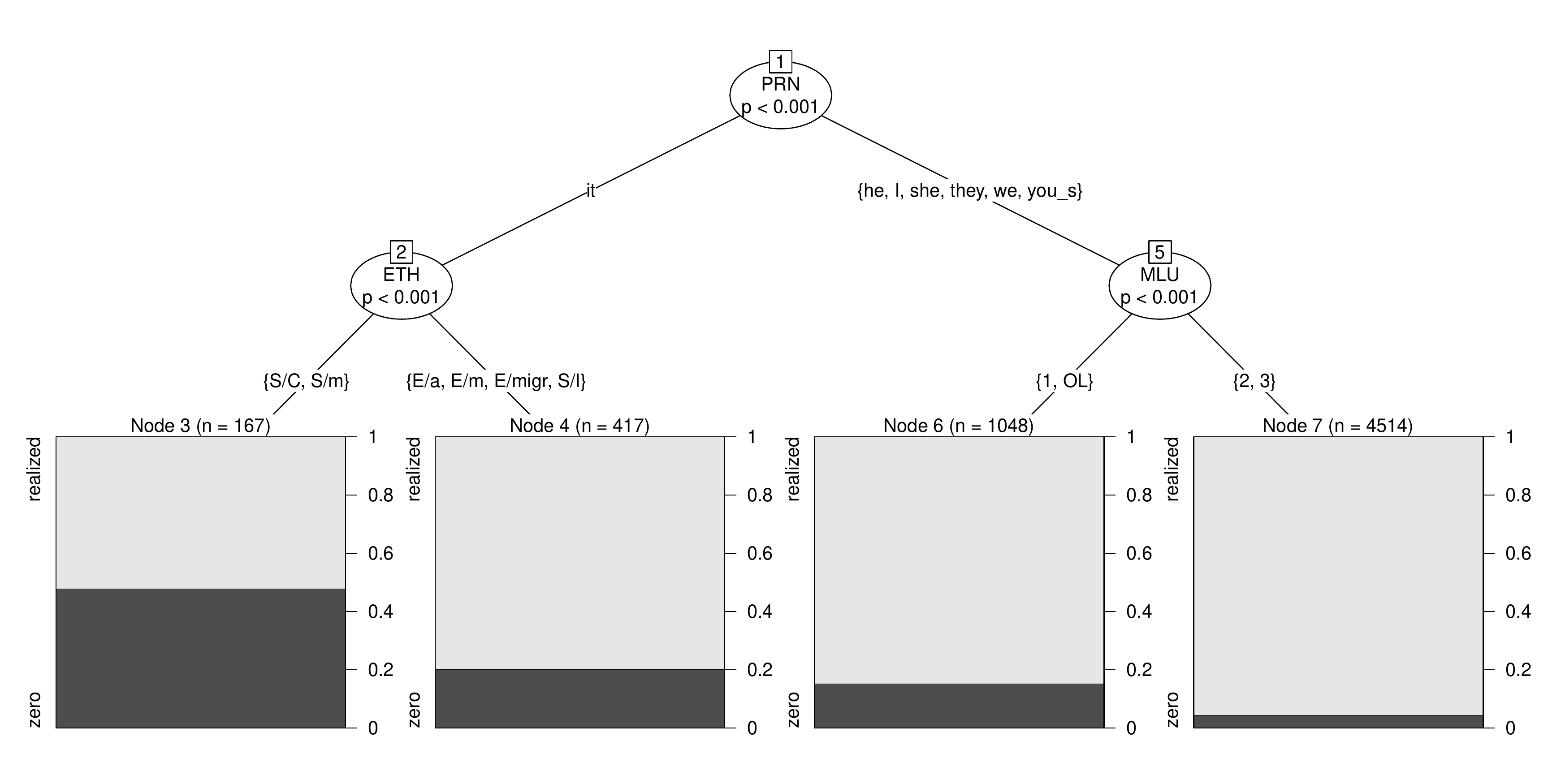}\\
\vspace{-0.3cm}
\caption{Simple decision tree with nearly identical class frequencies in one node.}
\label{fig:1}
\end{figure}

By default, ctrees predict the most frequent class in a terminal node. This way, the threshold for predicting, e.g., the smaller class is set to 0.5.
As illustrated by one of the trees in the PrInDT paper, sometimes one of the classes is predicted even though the two classes come with roughly the same frequencies (Node 3, Figure~\ref{fig:1}; taken from Weihs \& Buschfeld 2021a).

For dealing with extremely unbalanced class frequencies in random forests, Liaw \& Wiener (2002) suggest to alter the prediction rule to other than majority voting.
In the present paper, we aim to investigate whether the predictive power, represented by the balanced accuracy, is affected if we alter the prediction rule, i.e. the threshold at which the classes are selected for prediction. In particular, we assess whether lowering the threshold from which the smaller class is predicted affects the predictive power of the model. Such a threshold can be specified in the PrInDT extension RePrInDT.

\subsection{Variable importance in PrInDT ensembles}\label{subsec:4.3}

In linguistic studies, random forests are often used to complement ctree analyses to rank the predictors in their importance. Traditional random forests, however, cannot be used in PrInDT since they are based on resampling the same percentages of the small and the large classes. The unsuitability of random forests for modeling unbalanced classes is also reported in earlier linguistic studies (e.g. Lange \& Leuckert, 2021). RePrInDT, on the other hand, allows for different percentages of the large and the small classes, which is more adequate for modeling unbalanced data sets. In addition to that, traditional random forests use standard accuracies on test sets. PrInDT uses balanced accuracies. Traditional random forests are thus not adequate for measuring variable importance in PrInDT. Therefore, we aim at developing an alternative procedure.

How to measure variable importance in random forests has been extensively discussed in statistics. Strobl (2008), for example, comes to the conclusion that one should use subsampling and not bootstrapping for generating the data basis for the individual trees in the forest. Furthermore, to measure the importance of a predictor in a tree, one should compute the difference of the accuracies of the tree when applied to the original observations and after permuting the values of this predictor. These differences we call {\bf permutation losses}. Through the permutation, the order of the values of one of the predictors is manipulated in order to simulate the elimination of the predictor. This way, the effect of the elimination of the predictor is quantified. This procedure is repeated for all predictors and all trees and the accuracy differences for each predictor are summed up over the trees. Finally, the sums are normed by dividing each sum by the largest sum. This way, the most important predictor takes a relative importance of 100\%. The importance of the other predictors is represented as a percentage relative to the most important one. We integrate this approach in PrInDT by using the undersampling method (as introduced in Section~\ref{subsec:4.1}) for subsampling and by computing balanced accuracies for determining the permutation losses.

\section{Application to the linguistic example from PrInDT}\label{sec:5}

\subsection{Undersampling}\label{subsec:5.1}

In order to demonstrate the effect of varying the undersampling percentage of both the larger and the smaller class on the balanced accuracy, we use all possible combinations of psmall = {0.85, 0.90, 0.95, 1.0} and plarge = {0.07, 0.08, 0.09, 0.1} for our linguistic example. In order to guarantee identical starting points for random number generation, the same seed for the random numbers is used for each combination of psmall and plarge. As a test case for stratification, we use the predictor SEX, which is characterized by unbalanced level values. It comprises tokens from 4379 female and 1767 male children. Tables~\ref{tab:1} and \ref{tab:2} illustrate the results from the different undersampling combinations of plarge and psmall both for unstratified and stratified sampling. For the case of unstratified sampling, we take random undersamples from the two classes according to each of the specified combinations of percentages. In the case of proportional stratification, we take undersamples so that each class is undersampled in a way that the percentages of the levels of the predictor SEX from the original data set are retained. In the case of minimum criterion stratification, we restrict randomness in undersampling so that each value of the predictor SEX is represented by at least 340 tokens.

Table~\ref{tab:1} illustrates the results of the best trees of the 16 possible combinations of plarge and psmall. Table~\ref{tab:2} illustrates the results of the 16 ensembles of the 3 best trees. In both tables, columns 3--5 summarize the results from the unstratified runs, columns 6 and 7 provide the results from the stratified runs.

\begin{table}[H]
\caption{Accuracy of best trees for combinations of undersampling percentages}
\label{tab:1}
\begin{small}
\begin{tabular}{cc|ccc|cc}
& & \multicolumn{3}{c|}{unstratified} & prop. & min.crit.=340\\
& & \multicolumn{2}{c}{accuracy of} & balanced & strat. by SEX & strat. by SEX\\
plarge & psmall & realized & zero & accuracy & balanced acc. & balanced acc.\\
\toprule
 0.07  &   0.85  &   0.7823  &   0.6136  &   0.6980  &   0.7024  &   0.0 \\
 0.08  &   0.85  &   0.8083  &   0.5966  &   0.7024  &   0.6981  &   0.0 \\
 0.09  &   0.85  &   0.8104  &   0.5909  &   0.7007  &   0.7024  &   0.0 \\
 0.10  &   0.85  &   0.8090  &   0.5871  &   0.6981  &   0.6995  &   \textcolor{blue}{0.6986} \\
 0.07  &   0.90  &   0.7823  &   0.6136  &   0.6980  &   \textcolor{red}{\bf 0.7049}  &   0.0 \\
 0.08  &   0.90  &   0.8083  &   0.5966  &   0.7024  &   0.6980  &   0.0 \\
 0.09  &   0.90  &   0.8024  &   0.6042  &   \textcolor{red}{0.7033}  &   0.6990  &   0.0 \\
 0.10  &   0.90  &   0.8147  &   0.5833  &   0.6990  &   0.6990  &   0.6990 \\
 0.07  &   0.95  &   0.7823  &   0.6136  &   0.6980  &   0.7012  &   0.0 \\
 0.08  &   0.95  &   0.8024  &   0.6042  &   \textcolor{red}{0.7033}  &   0.7033  &   0.0 \\
 0.09  &   0.95  &   0.8024  &   0.6042  &   \textcolor{red}{0.7033}  &   0.7027  &   \textcolor{red}{0.7033} \\
 0.10  &   0.95  &   0.7898  &   0.6098  &   0.6998  &   0.7025  &   0.6998 \\
 0.07  &   1.00  &   0.8083  &   0.5966  &   0.7024  &   0.7024  &   0.0 \\
 0.08  &   1.00  &   0.7976  &   0.6080  &   0.7028  &   0.7024  &   0.7028 \\
 0.09  &   1.00  &   0.7981  &   0.6061  &   0.7021  &   0.7024  &   0.7021 \\
 0.10  &   1.00  &   0.8024  &   0.6042  &   \textcolor{red}{0.7033}  &   0.7024  &   \textcolor{red}{0.7033}
\end{tabular}
\end{small}
\end{table}

For the unstratified case, the best balanced accuracy of the {\bf best individual tree} for the different combinations of plarge and psmall is 0.7033 (marked in red in Table~\ref{tab:1}). This accuracy can be observed for four combinations of plarge and psmall (0.09/0.90, 0.08/0.95, 0.09/0.95, 0.10/1.00). The accuracies of the larger and smaller classes always range around 0.8 and 0.6, respectively. The balanced accuracy and thus the predictive power of the model does not strongly depend on the manifestations of psmall and plarge. Still, balanced accuracies are somewhat higher for plarge = 0.08 or 0.09 than for plarge = 0.07 or 0.1 for any value of psmall. In order to explain these results, we consider the sample sizes produced by undersampling at the different percentages. In our example, in order to generate a subsample of the larger class that has nearly the same size as the subsample of the smaller class,\\
\begin{small}
\hspace*{0.3cm} psmall * (no. of tokens in the smaller class) / (no. of tokens in the larger class)\\
\end{small}
needs to be between 0.08 and 0.093. This suggests that in our case undersampling should use plarge = 0.08 or 0.09 rather than plarge = 0.07 or 0.1. In addition to that, Table 1 shows that using the full small class for model construction does not lead to high `overfitting' since the balanced accuracies for psmall = 1 are in the same range as for the other choices of psmall.

The best balanced accuracy for the ensembles of the 3 best trees in the unstratified case is 0.7033 again (see Table~\ref{tab:2}). This time, this is attained only for psmall = 0.95 and plarge = 0.08. Accuracies of the larger and smaller class are, again, always around 0.8 and 0.6, respectively. Balanced accuracies for the ensembles are often slightly lower than for the best individual trees.

\begin{table}[H]
\caption{Accuracies of ensembles of the 3 best trees for combinations of undersampling percentages}
\label{tab:2}
\begin{small}
\begin{tabular}{cc|ccc|cc}
& & \multicolumn{3}{c|}{unstratified} & prop. & min.crit.=340\\
& & \multicolumn{2}{c}{accuracy of} & balanced & strat. by SEX & strat. by SEX\\
plarge & psmall & realized & zero & accuracy & balanced acc. & balanced acc.\\
\toprule
 0.07  &   0.85  &   0.7823  &   0.6136  &   0.6980  &   0.7024  &   0.0 \\
 0.08  &   0.85  &   0.7981  &   0.6061  &   0.7021  &   0.6980  &   0.0 \\
 0.09  &   0.85  &   0.8101  &   0.5871  &   0.6986  &   0.7024  &   0.0 \\
 0.10  &   0.85  &   0.7823  &   0.6136  &   0.6980  &   0.7023  &   0.6980 \\
 0.07  &   0.90  &   0.7823  &   0.6136  &   0.6980  &   0.6980  &   0.0 \\
 0.08  &   0.90  &   0.7823  &   0.6136  &   0.6980  &   0.6980  &   0.0 \\
 0.09  &   0.90  &   0.8054  &   0.6004  &   0.7029  &   0.6980  &   0.0 \\
 0.10  &   0.90  &   0.7823  &   0.6136  &   0.6980  &   0.7024  &   0.6980 \\
 0.07  &   0.95  &   0.7823  &   0.6136  &   0.6980  &   0.6859  &   0.0 \\
 0.08  &   0.95  &   0.8024  &   0.6042  &   \textcolor{red}{0.7033}  &   \textcolor{red}{0.7024}  &   0.0 \\
 0.09  &   0.95  &   0.8101  &   0.5871  &   0.6986  &   0.7020  &   0.6986 \\
 0.10  &   0.95  &   0.8101  &   0.5871  &   0.6986  &   0.6980  &   0.6986 \\
 0.07  &   1.00  &   0.8083  &   0.5966  &   0.7024  &   0.7024  &   0.0 \\
 0.08  &   1.00  &   0.8083  &   0.5966  &   0.7024  &   0.7024  &   \textcolor{blue}{\bf 0.7044} \\
 0.09  &   1.00  &   0.7981  &   0.6136  &   0.6995  &   0.7024  &   0.6995 \\
 0.10  &   1.00  &   0.8054  &   0.6004  &   0.7029  &   0.6980  &   \textcolor{blue}{0.7024}
\end{tabular}
\end{small}
\end{table}

In the case of proportional stratification, the results are similar. However, the best balanced accuracy of the best trees is slightly higher (0.7049) than in the unstratified case (0.7033) (see Table~\ref{tab:1}). In contrast, for the ensembles, the best balanced accuracy is slightly lower in the stratified case than in the unstratified case (see Table~\ref{tab:2}). In our example, proportional stratification by SEX does not improve the results.

If we set the minimum number of tokens for each category of the predictor SEX to 340 (min.crit.strat.), this requirement cannot be met in 9 of the 16 cases. This is indicated by zero (0.0) in Tables 1 and 2. In most of the cases where the requirement is met, the balanced accuracies equal those of the unstratified case. In those cases where sampling is repeated to meet the requirement (see Section 4.1), balanced accuracies may vary from the unstratified case (see the blue values in Tables~\ref{tab:1} and \ref{tab:2}). The best balanced accuracy overall for this kind of stratification is 0.7044 for one of the ensembles of 3 best trees. In general, this kind of stratification does also not improve the results.

\subsection{Prediction threshold}\label{subsec:5.2}

In a next step, we alter the threshold of the token frequencies in the terminal nodes of ctrees from which on the smaller class is predicted (see Section 4.2). We employ the thresholds of 0.5 (default), 0.45, 0.4, 0.35, 0.3, 0.25, and 0.2 on the same undersampling percentages as in the PrInDT paper, i.e. psmall = 1 and plarge = 0.09. As Table~\ref{tab:3} illustrates, the balanced accuracy of the best tree from undersampling decreases with the threshold. The balanced accuracy of the tree based on all observations, however, profits from lowering the threshold, in particular if the threshold is strongly decreased (to 0.2 in our example, see Table~\ref{tab:3}).

Therefore, as already suggested by Liaw \& Wiener (2002), lowering the threshold could be an alternative to undersampling for leveling out the imbalance between classes. Both methods increase the weight and influence of the smaller class in the model, but in different ways. Undersampling improves the balance between the smaller and the larger class, whereas lowering the threshold enhances the probability of the smaller class of being selected for prediction. Note, however, that the balanced accuracy of trees based on all observations only reflects goodness of fit and not predictive power.

\vspace{-0.3cm}
\begin{table}[H]
\caption{Balanced accuracies for different thresholds and trees}
\label{tab:3}
\vspace{0.2cm}
\begin{small}
\begin{tabular}{l|ccccccc}
trees / threshold & 0.5 & 0.45 & 0.4 & 0.35 & 0.3 & 0.25 & 0.2\\
\toprule
best undersampled  &  0.702 & 0.702 & 0.702 & 0.702 & 0.691 & 0.568 & 0.560\\
on all observations  &  0.500 & 0.568 & 0.568 & 0.568 & 0.568 & 0.568 & 0.671\\
\end{tabular}
\end{small}
\end{table}

\subsection{Ranking of predictors}\label{subsec:5.3}

\begin{figure}[H]
\centering
\vspace{-0.9cm}
\includegraphics[width=0.7\textwidth]{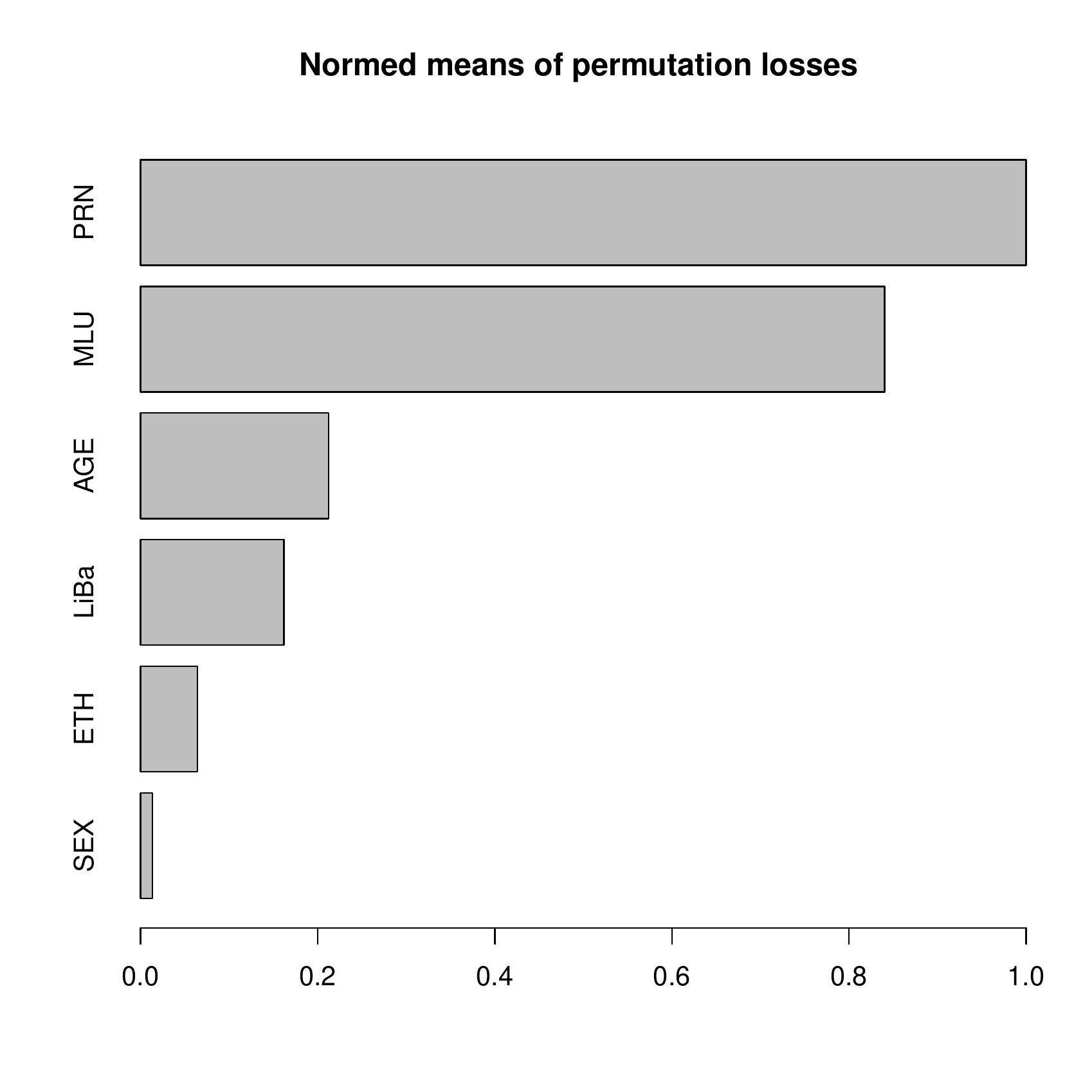}\\
\vspace{-0.3cm}
\caption{Normed means of permutation losses in the unstratified case.}
\label{fig:2}
\end{figure}

\begin{figure}[H]
\centering
\vspace{-0.9cm}
\includegraphics[width=0.7\textwidth]{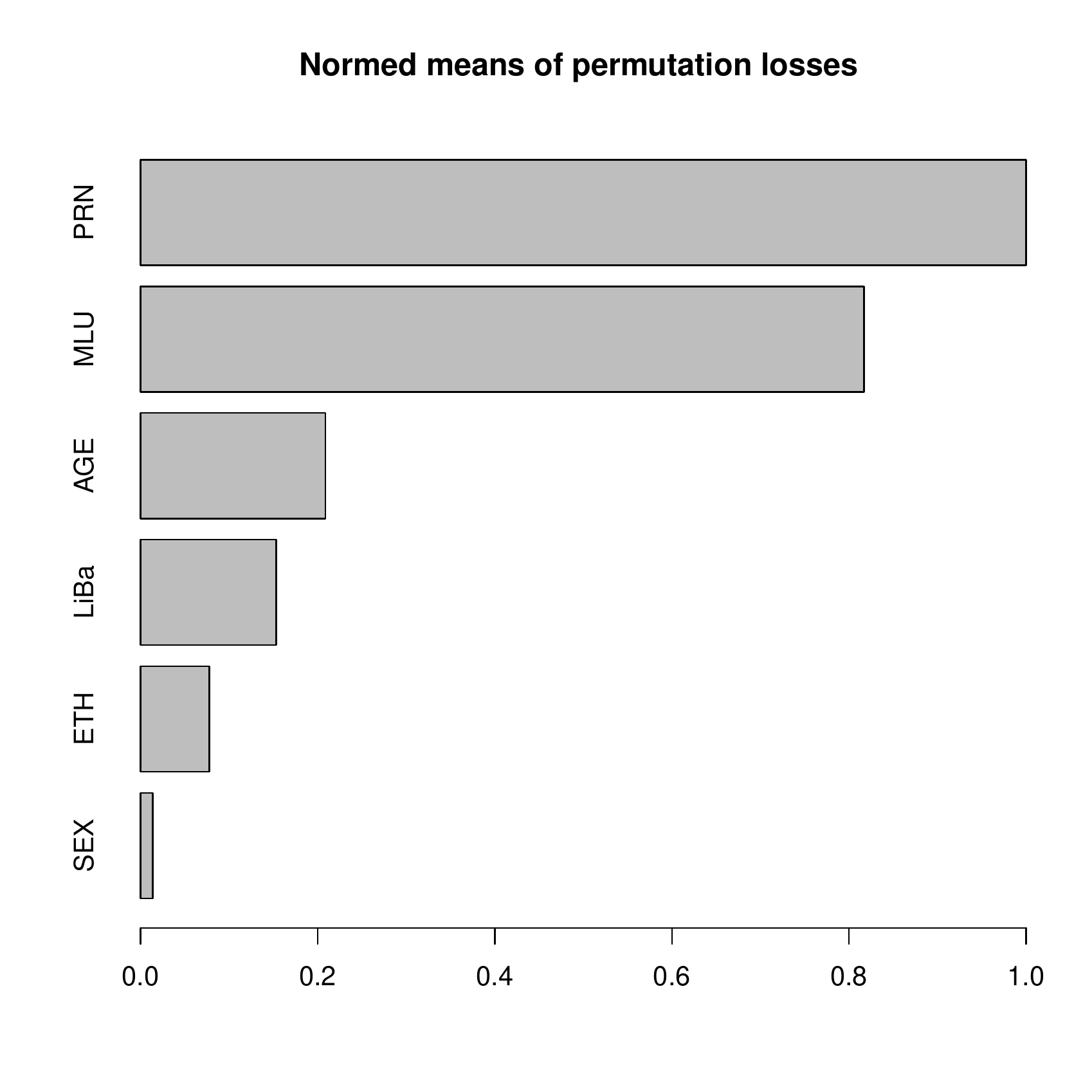}\\
\vspace{-0.3cm}
\caption{Normed means of permutation losses in case of proportional stratification for the predictor SEX.}
\label{fig:3}
\end{figure}

\begin{figure}[H]
\centering
\vspace{-0.9cm}
\includegraphics[width=0.7\textwidth]{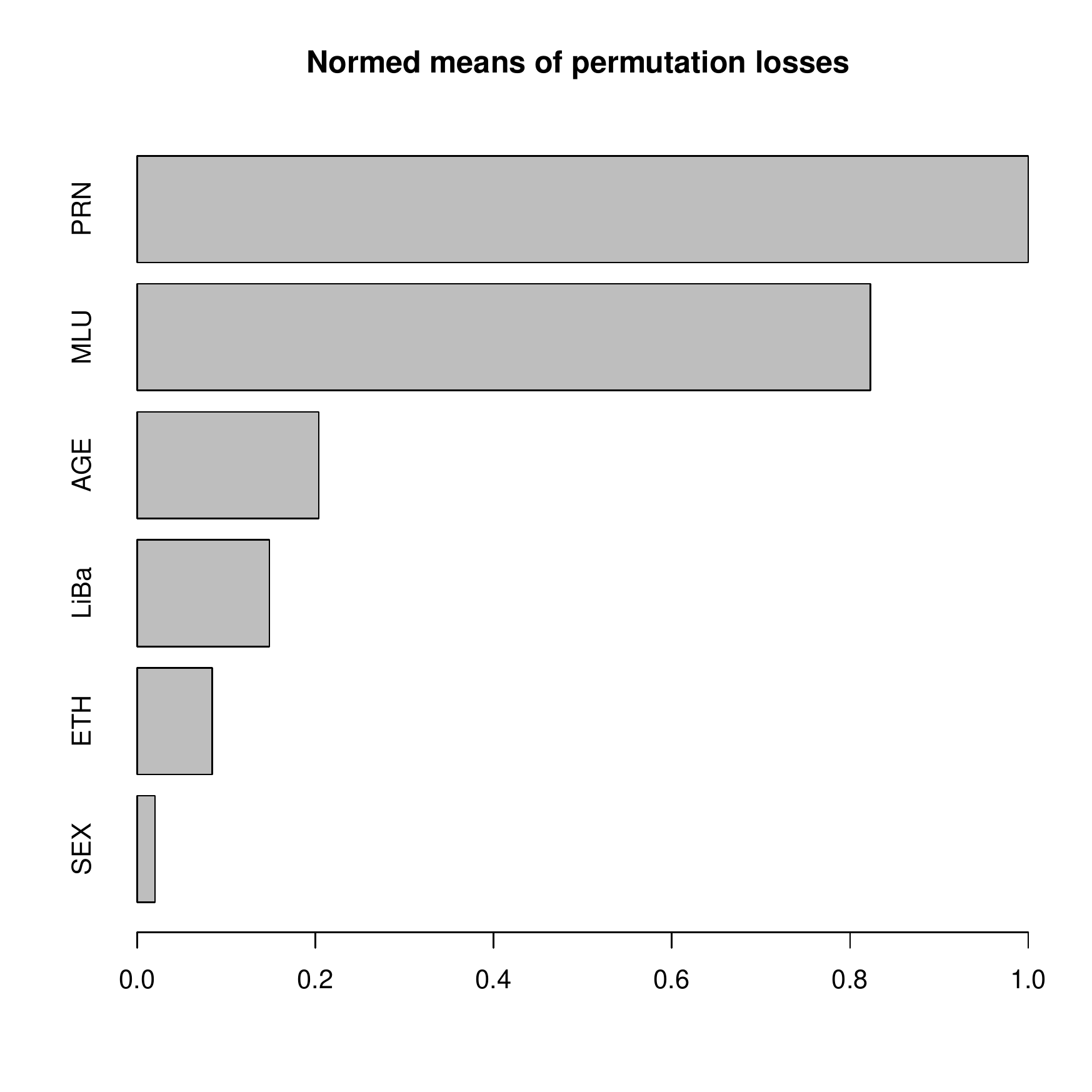}\\
\vspace{-0.6cm}
\caption{Normed means of permutation losses in case of minimum criterion (340) stratification for the predictor SEX.}
\label{fig:4}
\end{figure}

In order to determine predictor importance, we apply the procedure of measuring permutation losses (see Section 4.3) to the ensemble of the three best trees of all 16 combinations of plarge and psmall identified in Section~\ref{subsec:5.1}, i.e. to an ensemble of 48 trees. We compare the results from unstratified undersampling with those from stratified undersampling. The results are presented in Table~\ref{tab:4} and Figures~\ref{fig:2} through \ref{fig:4}.

\vspace{-0.3cm}
\begin{table}[H]
\caption{Means of permutation losses for the predictors based on the 3 best trees of the 16 ensembles}
\label{tab:4}
\vspace{0.2cm}
\begin{small}
\begin{tabular}{l|ccccccc}
& AGE & LiBa & ETH & SEX & MLU & PRN \\
\toprule
unstratified & 0.0658 & 0.0549 & 0.0152 & 0.0047 & 0.2672 & 0.3144\\
prop. strat. & 0.0645 & 0.0517 & 0.0126 & 0.0055 & 0.2625 & 0.3153\\
min.crit.(=340)strat. & 0.0286 & 0.0208 & 0.0118 & 0.0028 & 0.1153 & 0.1402\\
\end{tabular}
\end{small}
\end{table}

For the unstratified case, the ranking according to permutation losses identifies PRN as the most important predictor and MLU as the second most important predictor (Figure~\ref{fig:2}). AGE, LiBa, ETH, and SEX appear to be less important. Figures~\ref{fig:3} through \ref{fig:4} show similar results for the stratified cases. Nevertheless, Table~\ref{tab:4} shows much lower means of permutation losses when the min.crit.strat.\! is applied. This is because the min.crit.\! is not met in 9 of 16 cases and for such cases zero permutation losses are assumed. Still, Figure~\ref{fig:4} is similar to Figures~\ref{fig:2} and \ref{fig:3}, because values are normalized. Furthermore, the importance of SEX is not increased by stratifying this predictor.

\section{Discussion and conclusion}\label{sec:6}

In this paper, we have developed an extended version of PrInDT (Weihs \& Buschfeld, 2021a). First of all, we have introduced undersampling alternatives for leveling unbalanced classes. We have discussed the effects of varying the undersampling percentages plarge and psmall and of proportional and minimum criterion stratification. The results have revealed that the two stratification strategies do not lead to enhanced accuracies and suggest that the data should be undersampled in such a way that the choice of the percentages plarge and psmall generates subsamples of the classes of a similar size.

In a next step, we outlined the effects of modifying the prediction threshold. We have seen that this does not have an adjuvant effect on the balanced accuracy for models based on the undersampled data, but only for models based on the full data set. Since we are interested in prediction rather than model fit, this strategy is not convincing in the first place. However, the procedure needs to be tested on further data sets to assess its overall power.

In general, we have seen that both, lowering the threshold and undersampling of the larger class, can increase the weight and influence of the smaller class in the model. However, using both methods might overemphasize the smaller class leading to good predictions for the smaller class, but worse predictions for the larger one, and therefore lower balanced accuracies. Therefore, the joint use of the two methods needs to be carefully considered when it comes to their effect on the balanced accuracy.

Finally, we have introduced an alternative method for determining variable importance. Traditional random forests cannot be used in PrInDT since they are based on resampling the same percentages of the small and the large classes. We therefore adapt the method employed in random forests and introduce the notion of normed permutation losses (see Section~\ref{subsec:4.3}). Through normalization, the relative importance of the predictors comes to the fore. This appears to be more straightforward to interpret than the unnormed measures in the random forests.

Still, the validity and scientific power of the suggested methods remain to be assessed by future research. We would like to encourage interested colleagues to validate our findings and methods in their future work. PrInDT and all its extensions are written in the software R (R-Core Team 2019). The extensions are included in the R-function RePrInDT available from the authors (\texttt{claus.weihs@tu-dortmund.de}).

\section{References}
\begin{hangparas}{.25in}{1}

Hothorn, T., Hornik, K., Zeileis, A. 2006. Unbiased recursive partitioning: a conditional inference framework. J. Comput. Graph. Stat. 15, 651--674.

Lange, C., Leuckert, S. 2021. Tag questions and gender in Indian English; in: Bernaisch, T. (ed.), Gender in World Englishes; Cambridge University Press, 69--93

Liaw, A., Wiener, M. 2002. Classification and Regression by
randomForest. R News 2/3, 18--22


%

R Core Team 2019. R: A language and environment for statistical computing. R Foundation for Statistical Computing. \url{https://www.R-project.org/}.

Strobl, C. 2008. Statistical issues in machine learning - Towards reliable split selection and variable importance measures. Dissertation, Universität München, Fakultät für Mathematik, Informatik und Statistik.

Weihs, C., Buschfeld, S. 2021a. Combining Prediction and Interpretation in Decision Trees (PrInDT) - a Linguistic Example. arXiv: \url{http://arxiv.org/abs/2103.02336}.

Weihs, C., Buschfeld, S. 2021b. NesPrInDT: Nested undersampling in PrInDT. arXiv: \url{https://arxiv.org/abs/2103.14931}.

\end{hangparas}

\end{document}